%% file: 202005ICASSP.tex
\documentclass{article}
\usepackage{graphicx}
\usepackage[driverfallback=dvipdfm]{hyperref}
\usepackage{spconf,amsmath}
\usepackage{amssymb}
\usepackage{cite}
\usepackage{bm}
\usepackage{subcaption}
\usepackage{url}

\include{mathdef}

\newcommand{\sampleurl}{https://hyama5.github.io/demo_SRU_DGP_TTS}

\title{UTTERANCE-LEVEL SEQUENTIAL MODELING FOR DEEP GAUSSIAN PROCESS BASED SPEECH SYNTHESIS
USING SIMPLE RECURRENT UNIT}
\name{Tomoki Koriyama, Hiroshi Saruwatari\thanks{This work was partially supported by KAKENHI Grant Numbers 17K12711 and 19K20292.}}
\address{Graduate School of Information Science and Technology, The University of Tokyo, Japan \\
{\small \tt tomoki\_koriyama@ipc.i.u-tokyo.ac.jp}}

\begin{document}
\ninept
\maketitle
\begin{abstract}
This paper presents a deep Gaussian process (DGP) model with
a recurrent architecture
for speech sequence modeling.
DGP is a Bayesian deep model
that can be trained effectively with the consideration of model complexity
and is a kernel regression model that can have high expressibility.
In the previous studies, it was shown
that the DGP-based speech synthesis outperformed neural network-based one,
in which both models used a feed-forward architecture.
To improve the naturalness of synthetic speech, in this paper,
we show that DGP can be applied to utterance-level modeling
using recurrent architecture models.
We adopt a simple recurrent unit (SRU) for the proposed model to achieve
a recurrent architecture,
in which we can execute fast speech parameter generation
by using the high parallelization nature of SRU.
The objective and subjective evaluation results show that the proposed
SRU-DGP-based speech synthesis outperforms not only feed-forward DGP
but also automatically tuned SRU- and long short-term memory (LSTM)-based neural networks.
\end{abstract}
\begin{keywords}
speech synthesis, deep Gaussian process, simple recurrent unit,
Bayesian deep model, sequential modeling
\end{keywords}
%

\section{Introduction}

The quality of synthetic speech has largely been enhanced
by the advances of the deep architecture models using neural networks (NNs).
For example, it is reported that
the use of a huge amount of reading-style speech data of English could synthesize
speech that could not be distinguished from natural recordings \cite{shen2018natural,investigating2018luong}.
However, it is not easy to collect such a large amount of speech data
considering the use of various speech data
such as  minor languages, dialects, and conversational utterances.
Therefore, the choice of deep neural networks (DNNs) is not always appropriate
for arbitrary speech data
because the performance of NNs trained by a small amount of data
tends to be sensitive to model architectures and hyperparameters.

In this context, a speech synthesis framework based on a deep Gaussian process (DGP) \cite{damianou2013deep,salimbeni2017doubly}
was proposed as an alternative deep architecture model \cite{koriyama2019statistical}.
DGP is composed of multiple layers of GP regressions,
which are known as Bayesian kernel regressions.
Since DGP is a Bayesian framework,
we can train models while considering their complexity.
Moreover, DGP parameters are more interpretable than the weight matrices of NNs
because they are the representative pairs of inputs and outputs
of corresponding layers, called inducing points.
In the previous work \cite{koriyama2019statistical},
it was shown that DGP-based speech synthesis can generate more natural-sounding
speech than NN-based speech synthesis 
with well-tuned hyperparameters \cite{koriyama2019statistical}.
Furthermore,
we showed that the Bayesian nature of DGP easily leads to the
latent variable modeling of unannotated linguistic features \cite{koriyama2019semisupervised}.

Although the potential of DGP has been shown in the previous studies,
the evaluation in \cite{koriyama2019statistical} employed
simple frame-level modeling using feed-forward architectures.
On the other hand, NN-based speech synthesis studies
have shown the effectiveness of utterance-level modeling using
recurrent NNs (RNNs) and attention-based networks
\cite{fernandez2014prosody,wang2017tacotron}.
Therefore, we can expect that
utterance-level modeling of DGP-based speech synthesis
will improve its performance.

In recent years, many studies have attempted to show
the extensibility of DGPs to overcome
the limitations of feed-forward architecture.
The deep convolution Gaussian process \cite{kumar2018deep,dutordoir2019translation} achieves
a similar architecture to convolution NNs.
A probabilistic recurrent state-space model (PR-SSM) \cite{doerr2018probabilistic} mimics the RNN architecture
by representing the state transition function by GP regression.
However, PR-SSM is not appropriate for speech synthesis
because it requires GP regression recursively for each time step, and consequently,
the training and generation processes become slow.

In this study, we propose the utterance-level modeling of DGP-based speech synthesis using a simple recurrent unit (SRU) \cite{lei2018simple}.
SRU enables fast recurrent computation by using CNN-like parallelization
and outperformed the long short-term memory (LSTM)-RNN in some natural language processing tasks \cite{lei2018simple}.
The proposed SRU-DGP uses GP regression, which can be computed in parallel,
as an alternative to affine transformations of the original SRU.
Also, we reformulate the target function of DGP for utterance-level modeling
because the Monte Carlo sampling process in training should be performed
at an utterance level.
In experimental evaluations, we demonstrate the performance of the proposed SRU-DGP-based speech synthesis using various numbers of layers.
Subjective evaluation results indicate that the proposed technique
outperformed not only feed-forward DGP
but also SRU-NN and LSTM-RNN.
We also examine the difference in performance between DGP and Bayesian NN,
which are both Bayesian deep models.

\section{DGP-based Speech synthesis}
\label{sec:dgp}

The speech synthesis framework based on DGP proposed in \cite{koriyama2019statistical}
is based on the conventional DNN-based
framework \cite{zen2013statistical},
where we model the relationship between
a frame-level linguistic feature sequence $\XB$
and an acoustic feature sequence $\YB$
using deep architecture models.
The deep architecture models are expressed by
the composition of multiple functions as
\begin{align}
\YB &= f(\XB) + \epsilon \\
 f &= f^{L} \circ f^{L-1} \circ \dots \circ f^{1} 
\end{align} 
where $\epsilon$ is an element-wise noise.
General NN models use a combination of a linear (affine) transformation and
an activation function as a function $f^\ell$.
The key idea of DGPs is that
a latent function $f^\ell$ assumed to be distributed over a Gaussian process
and the posterior of the function is used for inference \cite{damianou2013deep}.

In DGP-based speech synthesis,
the training and inference are based on the technique of stochastic variational inference \cite{salimbeni2017doubly},
which is available for a large amount of training data.
Let $D_\ell$ be the dimensionality of layer $\ell$ and
$\HB^{\ell} = \left[\hB^{\ell,1}, \dots, \hB^{\ell,D_\ell}\right] = f^\ell(\dots(f^1(\XB)))$
be the hidden layer variables.
The relationship between the hidden layer variables sequence of $d$th dimension
$\hB^{\ell,d}$ and the lower layer variable $\HB^{\ell-1}$ is given by
the following posterior distribution:
\begin{align}
    q(\hB^{\ell, d}) &= \mathcal{N}\left(\hB^{\ell, d}; \muB^{\ell, d}, \SigmaB^{\ell, d}\right) \label{eq:svgp} \\
    \muB^{\ell, d} &= m(\HB^{\ell-1}) + \AB^\top \left(\mB^{\ell, d} - m(\ZB^{\ell})\right) \label{eq:svgp_mean} \\
    \SigmaB^{\ell, d} &= K(\HB^{\ell-1}, \HB^{\ell-1}) - \AB^{\top} \left(K(\ZB^{\ell}, \ZB^{\ell}) - \SB^{\ell, d}\right) \AB \\
    \AB &= K(\ZB^{\ell}, \ZB^{\ell})^{-1} K(\ZB^{\ell}, \HB^{\ell-1})
\end{align}
where $K(\cdot, \cdot)$ outputs a Gram matrix whose element is calculated by a positive definite kernel function $k(\xB, \xB')$.
The kernel function achieves the nonlinear mapping of the function $f^{\ell}$.
$m(\HB^{\ell-1})$ is the mean function of GP,
which is not generally used in GP regression.
$\mB^{\ell, d}$, $\SB^{\ell, d}$, and $\ZB^{\ell}$ are the parameters of the DGP model.
$\mB^{\ell, d}$ and $\SB^{\ell, d}$ denote
the mean and covariance of the variational distribution of inducing outputs $\uB^{\ell, d}$,
and $\ZB^{\ell}$ is an inducing input, which expresses
representative points of hidden layer variables $\HB^{\ell-1}$.

The parameter optimization of the DGP-based synthesis is
based on stochastic gradient descent
in the same manner as that of NN-based synthesis.
The target function to be maximized is a lower bound of marginal likelihood
called the evidence lower bound (ELBO).
ELBO is given by the following equation:
\begin{align}
    \mathcal{L} &= \frac{1}{S} \sum_{s=1}^S \sum_{i=1}^N 
    \Biggl\{ \sum_{d=1}^D \mathbb{E}_{q\left(h_{i,s}^{L,d} |\hat{\hB}_{i,s}^{L-1}\right)}
    \left[ \log p\left(y_i^d | h_{i,s}^{L,d}\right) \right]
    \notag \\ &~~~~ ~~~~
    - \frac{S}{N} \sum_{\ell=1}^{L} \sum_{d=1}^{D_\ell} \mathrm{KL}( q(\uB^{\ell,d}) \| p(\uB^{\ell,d} | \ZB^\ell) ) \Biggr\}
    \label{eq:dsvi_elbo}
\end{align}
where $N$ is the number of frames of training data, and
$S$ is the number of Monte Carlo samplings
to obtain the hidden layer sample $\hat{\hB}_{i,s}^{L-1}$ of the $(L-1)$th layer.
The sampling process is carried out by repeating the inference of the predictive posterior of hidden layer variables using \ref{eq:svgp}
and Monte Carlo sampling from the posterior distribution; this is known as
the reparameterization trick
in the field of variational autoencoders \cite{kingma2013auto}.
Since the first and second terms of (\ref{eq:dsvi_elbo}) represent the data-fit and complexity penalty, respectively,
the model can be trained by considering the model complexity.

\section{Simple recurrent unit}


An issue of RNNs including LSTM is slow computation because
the affine transformations for each time step cannot be performed in parallel.
A simple recurrent unit (SRU) \cite{lei2018simple} is proposed as a simple network
in which the affine transformations can be computed simultaneously.
The SRU of the $\ell$th layer is defined by the transformation of an input layer sequence
$\HB^{\ell-1} = \left[\hB_1^{\ell-1}, \dots, \hB_T^{\ell-1}\right]^\top$
into an output layer sequence $\HB^{\ell} = \left[\hB_1^{\ell}, \dots, \hB_T^{\ell}\right]^\top$
as:
\begin{align}
\phiB_t^\ell &= \sigma \left(\WB_\phi^\ell \hB_t^{\ell-1} + \bB_\phi^\ell + \vB_\phi^\ell \odot \cB_{t-1}^\ell\right) \label{eq:sru1} \\
\cB_t^\ell &= \phiB_t^\ell \odot \cB_{t-1}^\ell + \left(1 - \phiB_t^\ell\right) \odot \left(\WB_c^\ell \hB_t^{\ell-1} + \bB_c^\ell\right) \label{eq:sru2} \\
\rB_t^\ell &= \sigma \left(\WB_r^\ell \hB_t^{\ell-1} + \bB_r^\ell + \vB_r^\ell \odot \cB_{t-1}^\ell\right) \label{eq:sru3} \\
\hB_t^\ell &= \rB_t^\ell \odot \cB_{t}^\ell + (1 - \rB_t^\ell) \odot \left(\WB_h^\ell \hB_t^{\ell-1} + \bB_h^\ell\right)  \label{eq:sru4}
\end{align}
where $\WB_\phi^\ell$, $\WB_c^\ell$, $\WB_r^\ell$, and $\WB_h^\ell$ are weight matrices
and $\bB_\phi^\ell$, $\bB_c^\ell$, $\bB_r^\ell$, $\bB_h^\ell$, $\vB_\phi^\ell$, and $\vB_r^\ell$ are parameter vectors.
The operators $\odot$ and $\sigma$ denote an element-wise product and a
sigmoid function, respectively.

The SRU consists of two components:
\textit{light recurrence} (Eqs. (\ref{eq:sru1}) and (\ref{eq:sru3}))
and \textit{highway network} (Eqs. (\ref{eq:sru2}) and (\ref{eq:sru4})).
The recurrent computation in SRU is carried out using a state $\cB_t^\ell$
similarly to that using the memory cell in LSTM.
The state $\cB_t^\ell$ is updated for each time step using a forget gate
$\phiB_t^\ell$.
The layer output $\hB_t^\ell$ is determined by the state $\cB_t^\ell$ and an output gate $\rB_t^\ell$.

In contrast to RNNs, which use the past output vector $\hB_{t-1}^{\ell}$ for affine transformation,
the affine transformations of SRU depend only on the input vector $\hB_t^{\ell-1}$.
Therefore, the calculation time of affine transformations can be reduced by using parallel computing such as with a GPU.
Regarding the recurrent computation,
the operations for the state $\cB_t^\ell$ in (\ref{eq:sru1}) to (\ref{eq:sru4})
are either element-wise multiplication or addition,
whose computational complexities are much smaller than those of affine transformations.
Although the expressiveness of SRU is lower than that of LSTM and gated recurrent unit (GRU),
it has been reported that
SRU outperformed them in some language processing tasks \cite{lei2018simple}.



\begin{figure}[t]
\centering
\includegraphics[width=0.95\hsize]{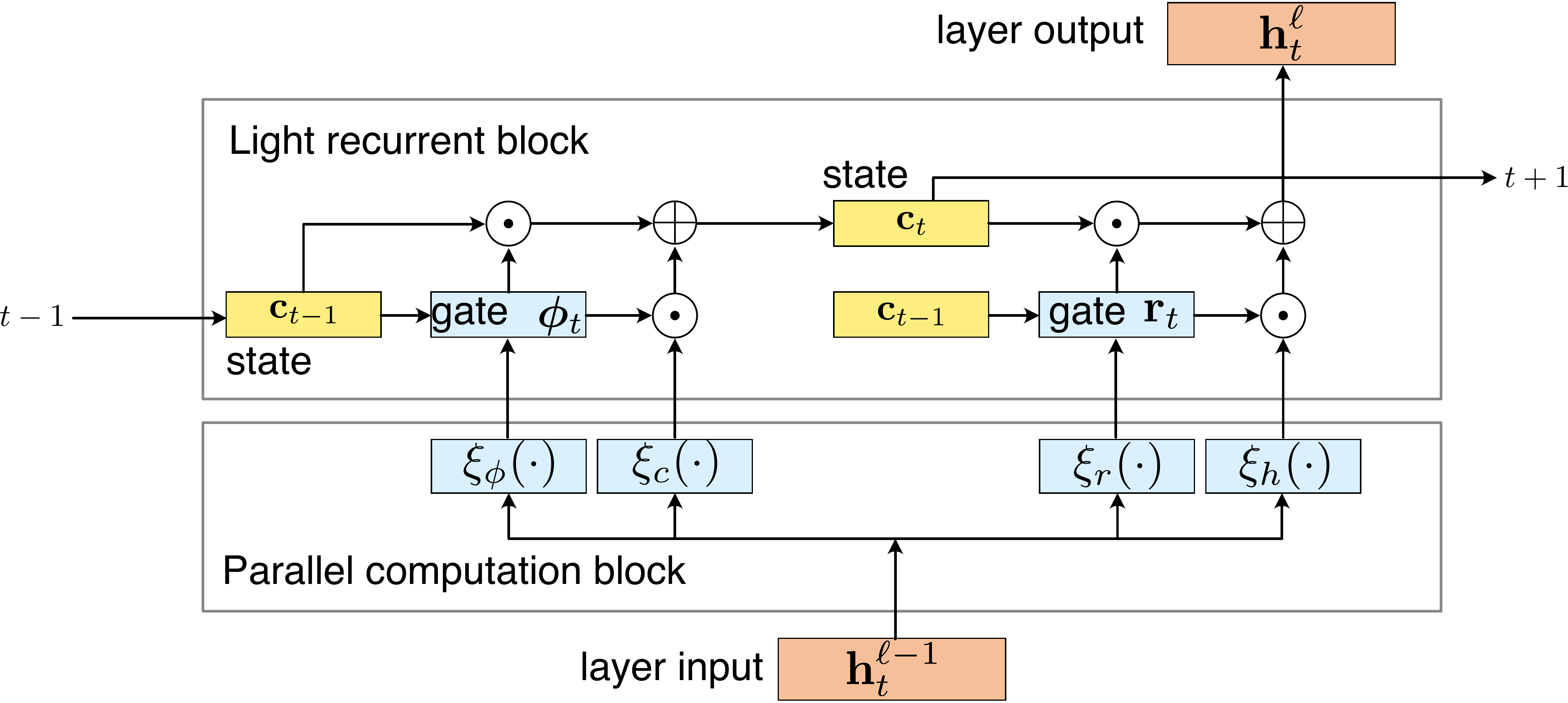}
\vspace{-5pt}
\caption{Flow of SRU-DGP block at time $t$ and layer $\ell$.}
\label{fig:sru_dgp}
\vspace{-5pt}
\end{figure}

\section{SRU for DGP-based speech synthesis}

\subsection{Generalization of SRU for DGP}

The key idea of DGP is to replace partial functions
with the functions over GPs.
We propose the SRU-DGP model by generalizing the affine transformations
in (\ref{eq:sru1}) to (\ref{eq:sru4}) to
latent functions, $\xi_\phi^\ell$, $\xi_c^\ell$, $\xi_r^\ell$, and , $\xi_h^\ell$,
which are distributed over GPs as
\begin{align}
\phiB_t^\ell &= \sigma \left(\xi_\phi^\ell(\hB_t^{\ell-1}) + \vB_\phi^\ell \odot \cB_{t-1}^\ell\right) \label{eq:sru_dgp1} \\
\cB_t^\ell &= \phiB_t^\ell \odot \cB_{t-1}^\ell + \left(1 - \phiB_t^\ell\right) \odot \xi_c^\ell(\hB_t^{\ell-1}) \label{eq:sru_dgp2} \\
\rB_t^\ell &= \sigma \left(\xi_r^\ell(\hB_t^{\ell-1}) + \vB_r^\ell \odot \cB_{t-1}^\ell\right) \label{eq:sru_dgp3} \\
\hB_t^\ell &= \rB_t^\ell \odot \cB_{t}^\ell + (1 - \rB_t^\ell) \odot \xi_h^\ell(\hB_t^{\ell-1}). \label{eq:sru_dgp4}
\end{align}
The flow of SRU-DGP is illustrated in Fig.\,\ref{fig:sru_dgp}.
As shown in the figure, we can separate the parallel and recurrent computations.
The inference in SRU-DGP layer is performed by applying the GP regression
for the functions $\xi_\phi^\ell$, $\xi_c^\ell$, $\xi_r^\ell$, and , $\xi_h^\ell$
for all frames in one sequence simultaneously.
After that, we carry out the recurrent computation of (\ref{eq:sru_dgp1}) to (\ref{eq:sru_dgp4}).
Hence, we do not need to perform GP regression recursively for each time step.

\subsection{Utterance-level model for SRU-DGP}

We reformulate the target function of DGP for
utterance-level modeling of speech synthesis.
Although Monte Carlo sampling is performed individually for each frame
in the feed-forward architecture of the previous work \cite{koriyama2019statistical},
we should perform utterance-level sampling
because each SRU layer is regarded as an utterance-level function.
The difference of utterance-level sampling from conventional frame-level sampling is
that we have to consider the influence of the covariance of frames
for the utterance-level sampling.

ELBO for the utterance-level modeling of DGP is defined by
\begin{align}
    \mathcal{L} &= \frac{1}{S} \sum_{s=1}^S \sum_{u=1}^{U}
    \Biggl\{ \sum_{d=1}^D \mathbb{E}_{q\left(\hB_{u,s}^{L,d} |\hat{\HB}_{u,s}^{L-1}\right)}
    \left[ \log p\left(\yB_u^d | \hB_{u,s}^{L,d}\right) \right]
    \notag \\ &~~~~ ~~~~
    - \frac{ST_u}{N} \sum_{\ell=1}^{L} \sum_{d=1}^{D_\ell} \mathrm{KL}( q(\uB^{\ell,d}) \| p(\uB^{\ell,d} | \ZB^\ell) ) \Biggr\}
    \label{eq:sru_dgp_elbo}
\end{align}
where $u$ is an utterance index,
and $U$ and $T_u$ are the number of utterances and the number of frames of the $u$th utterance, respectively.
$\yB_u^d$ and $\hB_{u,s}^{L,d}$ correspond to the true and predicted output sequence of the $d$th dimension, respectively.
In the same manner as in frame-level sampling, the utterance-level hidden layer values
$\hat{\HB}_{u,s}^{\ell} = \left[\hat{\hB}_{u,s}^{\ell, 1}, \dots, \hat{\hB}_{u,s}^{\ell, D_\ell} \right]$ ~~~ $(\ell = 1, \dots, L-1)$
are sampled using the predictive distribution obtained by (\ref{eq:svgp}) for each iteration of training.
Since the utterance-level predictive distribution is a multivariate Gaussian distribution of the form of
$\mathcal{N}(\hB; \muB, \SigmaB)$,
a sample sequence is obtained using the equation $\hat{\hB} = \muB + \LB \epsilonB$,
where $\LB$ is a matrix that satisfies $\LB \LB^\top = \SigmaB$
and $\epsilonB$ is a standard normal random vector.
Practically, the decomposition of $\SigmaB$ tends to be unstable during training.
Therefore, in this study, we approximate $\SigmaB$ using a low-rank matrix
based on random feature expansion \cite{rahimi2008random,cutajar2017random},
in which a kernel function is approximated by the inner product of finite-length vectors.

\section{Experimental evaluations}

\begin{figure}[t]
\centering
\includegraphics[width=0.85\hsize]{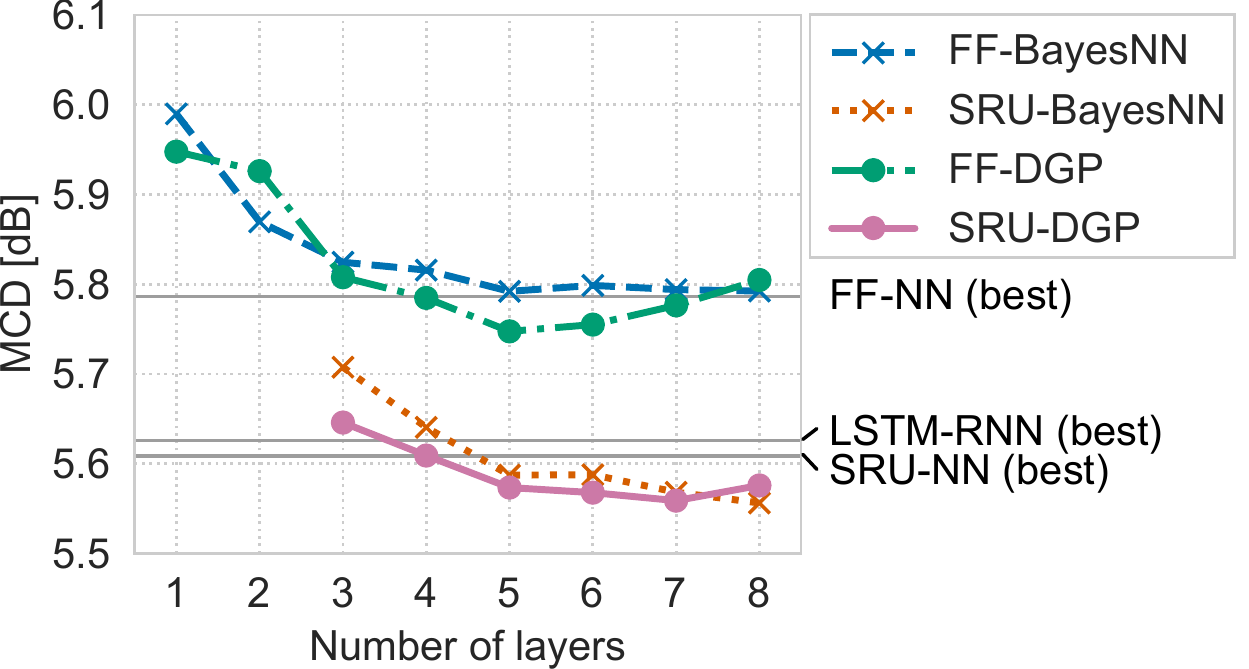}
\vspace{-5pt}
\caption{Spectral distortions between original and synthetic speech.}
\label{fig:mcd}
\vspace{5pt}
\centering
\includegraphics[width=0.85\hsize]{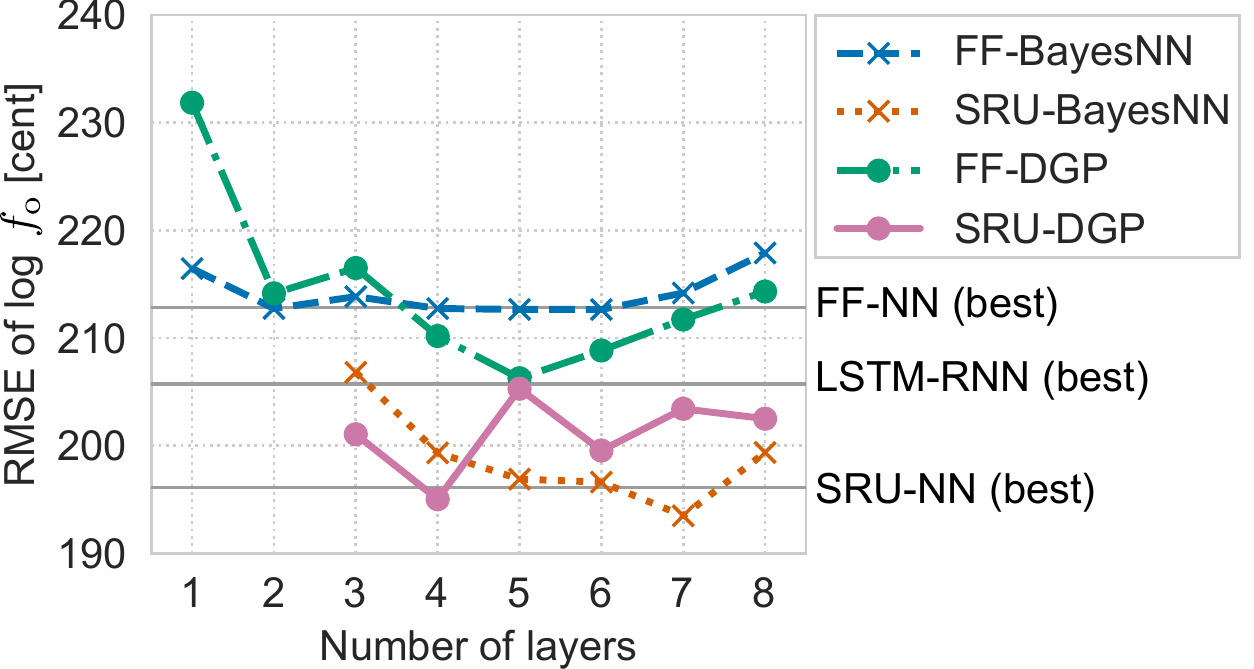}
\vspace{-5pt}
\caption{\fzero distortions between original and synthetic speech.}
\label{fig:lf0_rmse}
\vspace{-5pt}
\end{figure}

\subsection{Experimental conditions}

We used a Japanese speech database, JSUT corpus \cite{sonobe2017jsut},
for experimental evaluations
and chose the subset of BASIC0001 to BASIC2000.
OpenJTalk \cite{OpenJTalksite} was used to extract information about pronunciation and pitch accent.
The sentences with pronunciation extraction errors were removed.
We divided the data set into 1788, 60, and 60 sentences,
and used them as training, development, and test sets, respectively.
The amount of training data was approximately 1.95 h.
The acoustic features were extracted using the WORLD vocoder \cite{morise2016world} and SPTK \cite{SPTKsite}
from the speech samples that were down-sampled to 16 kHz.
We used 187-dimensional acoustic features as output features,
which consisted of the 0--59th mel-cepstrum, log \fzero, 1-dim code aperiodicity,
and their $\Delta$ and $\Delta^2$ features, and a voiced/unvoiced flag.
The input features were 575-dimensional context vectors of
binary-valued linguistic information and continuous-valued relative frame positions.

The model architectures of the proposed SRU-DGP method with $L ~(=3 \dots 8)$ layers
had feed-forward layers at the bottom and top layers.
We inserted $(L-2)$ SRU layers between the feed-forward layers.
The mean function had zero mean and
the kernel function was the ArcCos kernel \cite{cho2009kernel}
with normalization terms based on the results in \cite{koriyama2019statistical}.
The dimensionality of the hidden-layer variables was 256
and the number of inducing points of each layer was 1024.
To avoid the effect of initial values, we employed a layerwise pretraining technique \cite{koriyama2019training}
to determine the initial values of DGP parameters\footnote{We observed
that even random initialization achieved comparable or slightly larger acoustic feature distortions than those using the pretraining.}.
We fixed the parameters of SRU layers, $\vB_\phi$ and $\vB_r$,
to all-one vectors, because the optimization of the parameters tended to cause overfitting.
We used 1024-dimensional random feature expansion for utterance-level sampling.
The optimization of the model parameters was based on Adam \cite{kingma2014adam}
with a learning rate of $10^{-2}$, and one utterance was used as a minibatch.
We set the number of samples $S$ to unity.

We compared the proposed SRU-DGP with
not only conventional feed-forward DGP, but also
LSTM- and SRU-based NNs.
We performed hyperparameter tuning for the NNs
using Optuna \cite{akiba2019optuna}.
The hyperparameters were the numbers of layers and hidden units,
the learning rate of Adam, the dropout rate, the weight decay,
and the use of layer normalization and residual blocks.
We generated 100 candidates and chose the best hyperparameter set
that yielded the smallest validation loss.

Furthermore, we compared the proposed method with Bayesian NN \cite{pmlr-v32-rezende14},
which can be trained while considering the model complexity
in the same manner as DGP.
As Bayesian NN, we employed the training method based on stochastic variational inference \cite{hoffman2013stochastic}
similarly to the DGP-based speech synthesis.
The number of hidden units was 1024 and
Adam with a learning rate of $10^{-5}$ was used as an optimizer.
We used Gaussian distributions as the variational distributions of weight and bias parameters.
The mean vectors of the variational distribution were initialized randomly
and the initial variances were $10^{-4}$.
In the inference for Bayesian NN, we sampled 100 NNs and
used the mean of the output features for speech parameter generation.
The methods evaluated are summarized as:
\begin{description}
\item[FF-NN, SRU-NN, LSTM-RNN:] Conventional NN-based methods using feed-forward, SRU, and LSTM
blocks, respectively. We used the best hyperparameter sets, which were automatically tuned by Optuna, for the evaluations.
\item[FF-BayesNN, SRU-BayesNN:] Bayesian NN-based methods using feed-forward and SRU blocks.
\item[FF-DGP, SRU-DGP:] DGP-based methods. FF-DGP is the conventional method in \cite{koriyama2019statistical}. SRU-DGP is our proposed method.
\end{description}

The models of DGP, NN, and BayesNN were trained using PyTorch and
SRU implementation\footnote{\url{https://github.com/asappresearch/sru}}.
Computation times were evaluated using an NVIDIA TITAN Xp GPU and
an Intel Core i7-7700X CPU (4.2 GHz).


\subsection{Results}

\subsubsection{Objective evaluation}

To evaluate the model accuracy of the proposed method,
we calculated the acoustic feature distortions of mel-cepstrum and \fzero
between the original and synthetic speech.
Figures\,\ref{fig:mcd} and \ref{fig:lf0_rmse} respectively show the mel-cepstral distortions (MCDs) and root mean square errors (RMSEs) of log \fzero
as functions of the number of layers.
For NN-based methods, we calculated the distortions obtained by the best hyperparameter sets.
It can be seen that the proposed SRU-DGP gave significantly smaller distortion
than FF-DGP.
This indicates the effectiveness of the use of a recurrent architecture in DGP-based speech synthesis.
By comparing SRU-DGP with the NN methods of NN-SRU and LSTM-RNN,
the MCDs of the proposed SRU-DGP with more than four layers were found to be smaller than those obtained by the NN-based methods.
For the Bayesian methods, 
we see that SRU-DGP had slightly smaller MCDs than SRU-BayesNN
except for the eight-layer models.
On the other hand, the RMSEs of log \fzero of SRU-DGP were larger
than those of SRU-BayesNN when we use more than four layers.

\begin{figure}[t]
\centering
\includegraphics[width=0.95\hsize]{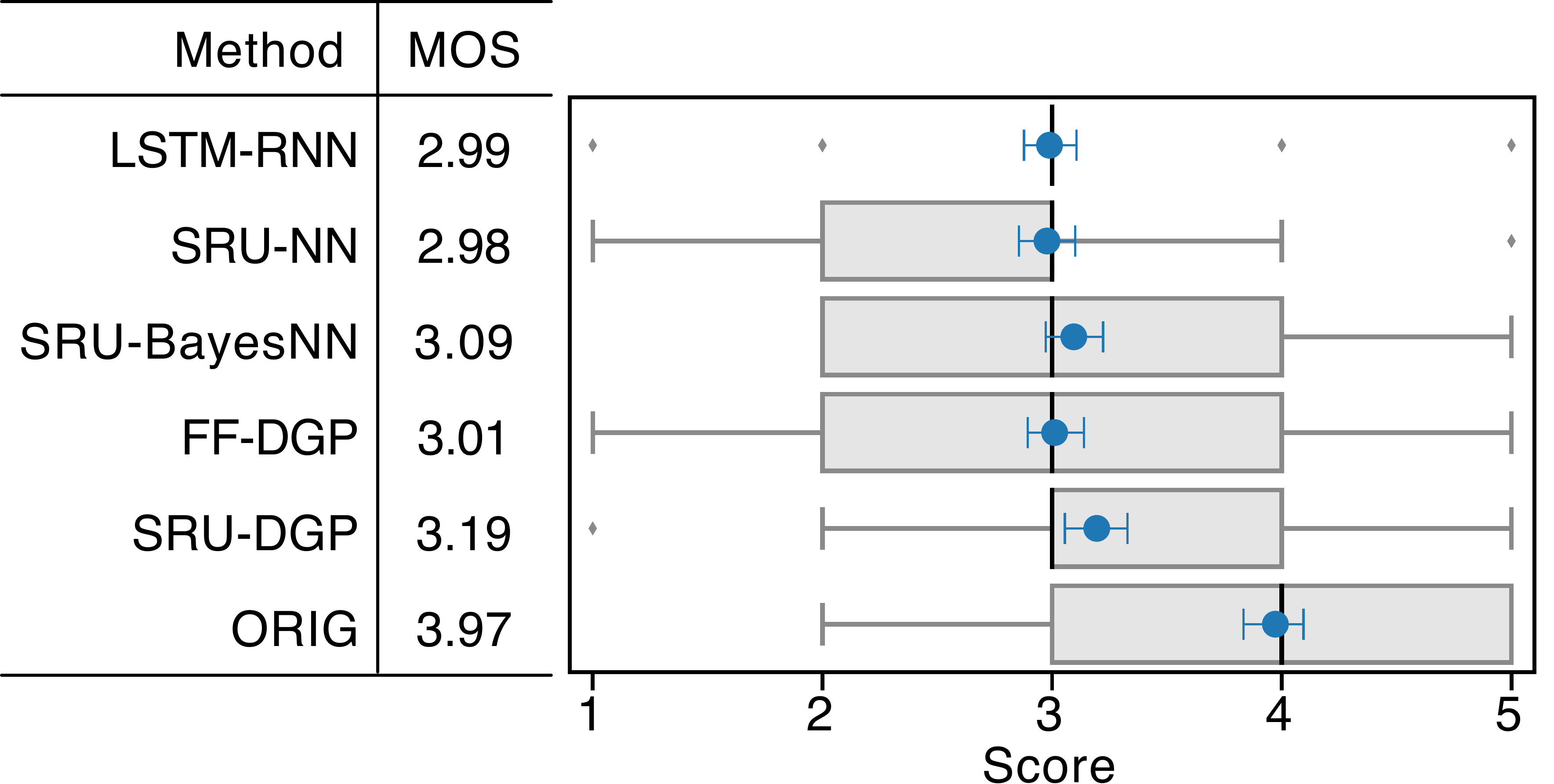}
\vspace{-5pt}
\caption{Subjective evaluation results on a mean opinion score (MOS) test.
Box plots represent the distributions of the scores. The blue-colored circles and
error bars are MOSs and 95\% confidence intervals, respectively.}
\label{fig:subj_mos}
\vspace{-5pt}
\end{figure}
 
\begin{figure}[t]
\centering
\includegraphics[width=0.85\hsize]{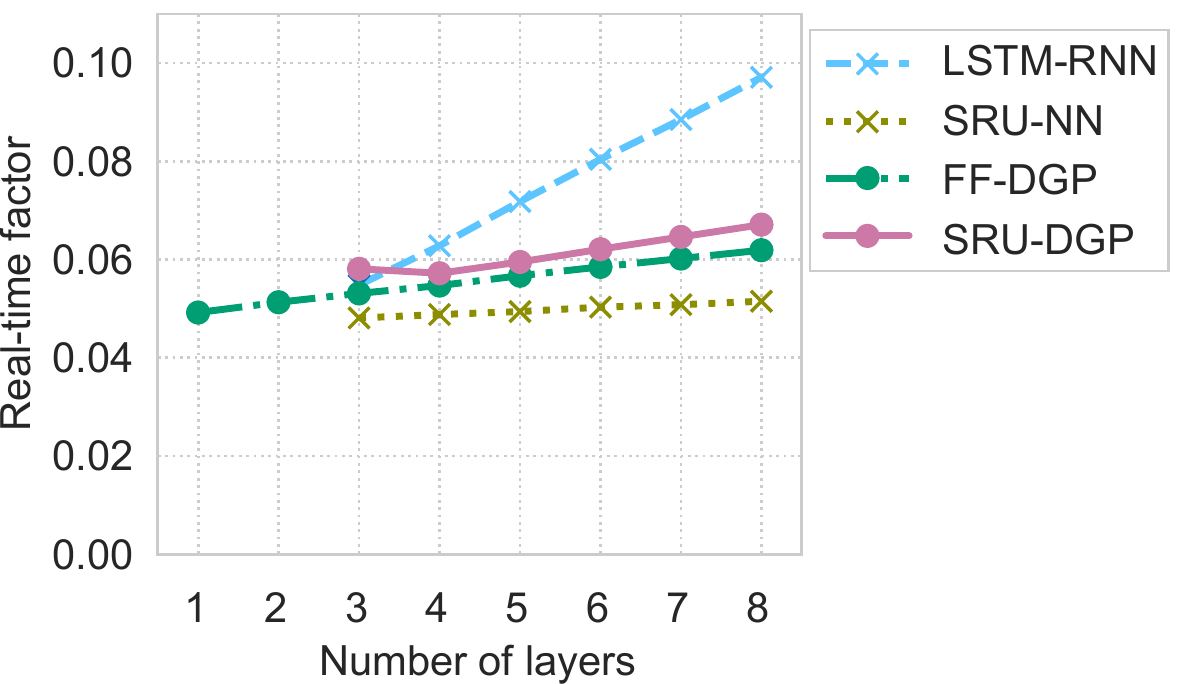}
\vspace{-5pt}
\caption{Real-time factors for the computation of speech parameter generation.}
\label{fig:exec_time}
\vspace{-5pt}
\end{figure}

\subsubsection{Subjective evaluation}

The perceptual quality of synthetic speech was evaluated by
a mean opinion score (MOS) test\footnote{Speech samples are available at \href{\sampleurl}{\url{\sampleurl}}}.
The best model was chosen by measuring validation loss
for SRU-BayesNN, FF-DGP, and SRU-DGP.
The subjective evaluation was carried out using
a crowd-sourcing service.
Sixty participants listened to 18 utterances (three sentences, six methods)
and rated the naturalness on a five-point scale: 5: excellent, 4: good, 3: fair, 2: poor, and 1: bad.

The results are shown in Fig.\,\ref{fig:subj_mos}.
ORIG in the figure represents the original recording down-sampled to 16 kHz.
According to the paired t-test with $\alpha=0.05$, 
the MOS of SRU-DGP was significantly higher than those of conventional LSTM-RNN, SRU-RNN, and FF-DGP.
We also note that SRU is effective for the speech synthesis task
because the scores of SRU-NN and LSTM-RNN were comparable.
Although there was no significant difference between SRU-DGP and SRU-BayesNN,
a score of 2 (poor) for SRU-DGP was rarely chosen.

\subsubsection{Execution time}

To evaluate the computational complexity of the proposed SRU-DGP models,
we measured the execution time on synthesizing speech.
The number of hidden units was set to 1024
for the conventional methods of SRU-NN and LSTM-RNN.
Figure\,\ref{fig:exec_time} shows the real-time factors (RTFs)
for speech parameter generation, which are averages of 10 trials.
The RTFs were calculated by dividing the average execution time
by the length of non-silence speech segments.
Note that the RTFs in Fig.,\ref{fig:sru_dgp} do not include
waveform generation using WORLD,
which took approximately 0.24 RTF.

It can be seen from the figure
that SRU-DGP was faster than LSTM-RNN
whereas it was slower SRU-NN.
Compared with FF-DGP,
SRU-DGP was just slightly slower than FF-DGP.
This indicates that SRU-DGP can be incorporated
into complicated architectures of DGP models
without worrying about computation time.

\section{Conclusions}

In this paper, we have proposed a DGP-based speech synthesis framework,
in which utterance-level modeling is available by using
an SRU-based recurrent architecture.
Specifically, we replace the parallel computation of the affine transformation in SRU
by the function over GP
and perform GP regression simultaneously for all frames in one utterance.
A Monte Carlo sampling process for the training of DGP
is calculated using an utterance level.
The experimental results show that the proposed SRU-DGP outperformed not only
feed-forward DGP but also SRU- and LSTM-based NNs whose hyperparameters were
automatically optimized.
Future work will extend the DGP-based speech synthesis to
a wider variety of architectures.
For example, since state-of-the-art speech synthesis employs
an attention-based sequence-to-sequence network \cite{wang2017tacotron},
we will examine the extensibility of DGP to the attention architecture.

\newpage

\bibliographystyle{IEEEbib}
\bibliography{refs}

\end{document}

%% file: mathdef.tex
\usepackage{bm}

\newcommand{\epsilonB}{\bm{\epsilon}}
\newcommand{\muB}{\bm{\mu}}
\newcommand{\SigmaB}{\bm{\Sigma}}

\newcommand{\phiB}{\bm{\phi}}

\newcommand{\AB}{\mathbf{A}}

\newcommand{\bB}{\mathbf{b}}
\newcommand{\cB}{\mathbf{c}}

\newcommand{\hB}{\mathbf{h}}
\newcommand{\HB}{\mathbf{H}}

\newcommand{\LB}{\mathbf{L}}

\newcommand{\mB}{\mathbf{m}}

\newcommand{\rB}{\mathbf{r}}
\newcommand{\SB}{\mathbf{S}}

\newcommand{\uB}{\mathbf{u}}
\newcommand{\vB}{\mathbf{v}}

\newcommand{\WB}{\mathbf{W}}
\newcommand{\xB}{\mathbf{x}}
\newcommand{\XB}{\mathbf{X}}
\newcommand{\yB}{\mathbf{y}}
\newcommand{\YB}{\mathbf{Y}}

\newcommand{\ZB}{\mathbf{Z}}

\newcommand{\fzero}{{$f_\mathrm{o}~$}}